# An Intelligent Innovation Dataset on Scientific Research Outcomes

Xinran Wu, Hui Zou, Yidan Xing, Jingjing Qu, Qiongxiu Li, Renxia Xue, Xiaoming Fu*

**ABSTRACT**

Various stakeholders, such as researchers, government agencies, businesses, and research laboratories require a large volume of reliable scientific research outcomes including research articles and patent data to support their work. These data are crucial for a variety of application, such as advancing scientific research, conducting business evaluations, and undertaking policy analysis. However, collecting such data is often a time-consuming and laborious task. Consequently, many users turn to using openly accessible data for their research. However, these existing open dataset releases typically suffer from lack of relationship between different data sources and a limited temporal coverage. To address this issue, we present a new open dataset, the Intelligent Innovation Dataset (IIDS), which comprises six interrelated datasets spanning nearly 120 years, encompassing paper information, paper citation relationships, patent details, patent legal statuses, and funding information. The extensive contextual and extensive temporal coverage of the IIDS dataset will provide researchers and practitioners and policy maker with comprehensive data support, enabling them to conduct in-depth scientific research and comprehensive data analyses.

Specifications Table

| | |
|---|---|
| Subject | Scientific research achievements and patents |
| Specific subject area | 27 disciplines (subject classification observation points based on Scopus) |
| Data format | SQL; corpus text and additional information stored in Json-lines. |
| Type of data | Tables |
| Data collection | It contains detailed information on papers, patents, fundings, and more spanning nearly 120 years, totaling about 735.1GB.<br>The papers section includes English-language articles formally published in journals indexed by established systems, alongside related conference data and citation information. It covers representative online databases such as Web of Science, Scopus, and Springer.<br>The patents section primarily derives from the European Patent Office's official patent data, covering the period from 1950 to the present(https://worldwide.espacenet.com/).<br>The funds section compiles research projects from major countries worldwide, with primary sources including the United States, China, Japan, and Canada. This includes projects funded by major funding agencies such as the U.S. National Science Foundation, National Natural Science Foundation of China and Social Science Fund of China. |
| Data source location | •Institution: Research Center for Social Intelligence at Fudan University, Boguan Innovation (Shanghai) Big Data Technology Co., Ltd., and Shanghai Artificial Intelligence Laboratory.<br>• City/Town/Region: Shanghai.<br>• Country: China |
| Data accessibility | Repository name: ZHICHUANGDATA<br>https://openxlab.org.cn/datasets/Gracie/ZHICHUANG-DATA<br>Six distinct tables have been made available, with links to the open dataset provided for direct access.<br>1. entity_paper<br>2. reference_citation_re<br>3. entity_fund_re<br>4. entity_fund_info<br>5. base_patent_detail<br>6. base_patent_law_status |

## 1 Value of the Data

The Intelligent Innovation Dataset(IIDS) has been jointly developed by the Research Center for Social Intelligence at Fudan University, Boguan Innovation (Shanghai) Big Data Technology Co., Ltd., and Shanghai Artificial Intelligence Laboratory.

The Intelligent Innovation Dataset emphasizes presenting the panorama of technological innovation through data, extensively covering research achievements and patent-related data. It currently mainly includes various types of academic achievement data such as academic papers, research projects, patents, and more. The current size of the database is 735.1GB, including over 92.31 million academic paper records and over 1.8 billion citation relationship records; over 30,000 research fund information records and over 61.28 million paper-fund relationship records; over 100 million patent records and over 110 million patent legal status-related data records.

The characteristics of the Intelligent Innovation Dataset are as follows:

(1) Wide coverage: The Intelligent Innovation Dataset covers four major categories - natural sciences, medicine, social sciences, and life sciences – encompassing 27 major disciplines and 334 minor disciplines, offering comprehensive and extensive content.

(2) Rich content: It includes a wide range of academic materials, encompassing not only journal articles but also conference papers. Furthermore, it provides access to a substantial collection of patent data from the European Patent Office. Patents are a form of intellectual property and a measure of innovation output, which cover a broad range of technologies and are a rich source of information [1]. And The European Patent Office has become the de facto standard for researchers working with patent data [2].

(3) Extended historical span: The Intelligent Innovation Dataset contains detailed information on papers, patents, etc., from different countries spanning nearly 120 years, providing users with a broad research perspective and historical data reference.

In summary, the Intelligent Innovation Dataset, with its extensive data coverage, integration of patent data, and the characteristic of open access, offers a convenient one-stop information service for academic research and further application. It is positioned to become a significant resource for researchers and practitioners worldwide engaged in related studies.

## 2 Background

Big data from research papers and patents is key to grasping innovation. Research paper offer insights into trends and notable contributors, while analyzing citations and content can uncover chances for further innovation. On the other hand, patents serve as an essential indicator of technological innovation, offering insights into the development of new technologies and their impact on industries and society [3]. Moreover, applying a variety of tools such as social network analysis and regression models, researchers can exploit the full patents dataset not only for investigating and modelling their global knowledge dynamics and structures, but also for forecasting their evolutionary advancements [4]. The analysis of patent data also allows researchers to uncover technological trends, competitive landscapes, and emerging areas of research [5].

By harnessing big data analytics to integrate insights from research papers and patents, researchers and practitioners can gain a comprehensive understanding of innovation dynamics and make informed decisions. This holistic approach to leveraging big data sources contributes to shaping future research agendas, fostering innovation, and driving societal progress [6].

However, one significant gap lies in the integration and interoperability of diverse datasets from different sources. Often, datasets are siloed and fragmented, making it challenging to perform comprehensive analyses that span multiple domains and disciplines [7]. Therefore, there is a pressing need to develop datasets with high-quality data integration to enabling researchers to seamlessly combine and analyze diverse datasets.

## 3 Data Description

Six tables are provided (see Fig. 1). The first table (entity_paper) contains journal paper data, including scholarly papers(mainly in English) formally published in journals indexed in reputable systems and conference data related to the papers. Unpublished articles, preprints, non-journal web articles, and papers with the publication status "Online" are not included. This data is primarily sourced from the representative Scopus database. The field 'eid' within the dataset is unique can be used to achieve cross-referencing among multiple tables in the dataset.

The second table (reference_citation_re) contains all the paper citation relationships from Table 1(entity_paper), divided into ref_id and cited_id (essentially both are eid). Here, ref_id represents the reference paper, while cited_id represents the cited paper. Cross-referencing between the tables can be done using ‘eid’.

The third table (entity_fund_re) contains funding information related to papers, which can be matched with the Table 1 (entity_paper) using ‘eid’, and with the Table 4 (entity_fund_info) using ‘fid’ (also known as 'findid').

The fourth table (entity_fund_info) contains research projects issued by major countries globally, primarily sourced from countries such as the United States, China, Japan, and Canada. This includes projects funded by major funding agencies like the U.S. National Science Foundation, the European Union, the National Natural Science Foundation of China and National Social Science Fund of China. Matching can be accomplished using "fundid" in Table 3 (entity_funds_re). It is noteworthy that "fundid" appears in JSON format within the funding_list column of Table 1 (entity_paper).

The fifth table (base_patent_detail) contains patent data categorized into valid patents and valuable patents for different model calculations. This data is primarily sourced from the official website of the European Patent Office, covering the period from 1950 to the present(https://worldwide.espacenet.com/).

The six table (base_patent_law_status) contains legal information for each stage of the patents. It can be matched with the Table 5 (base_patent_detail) using 'pn'.

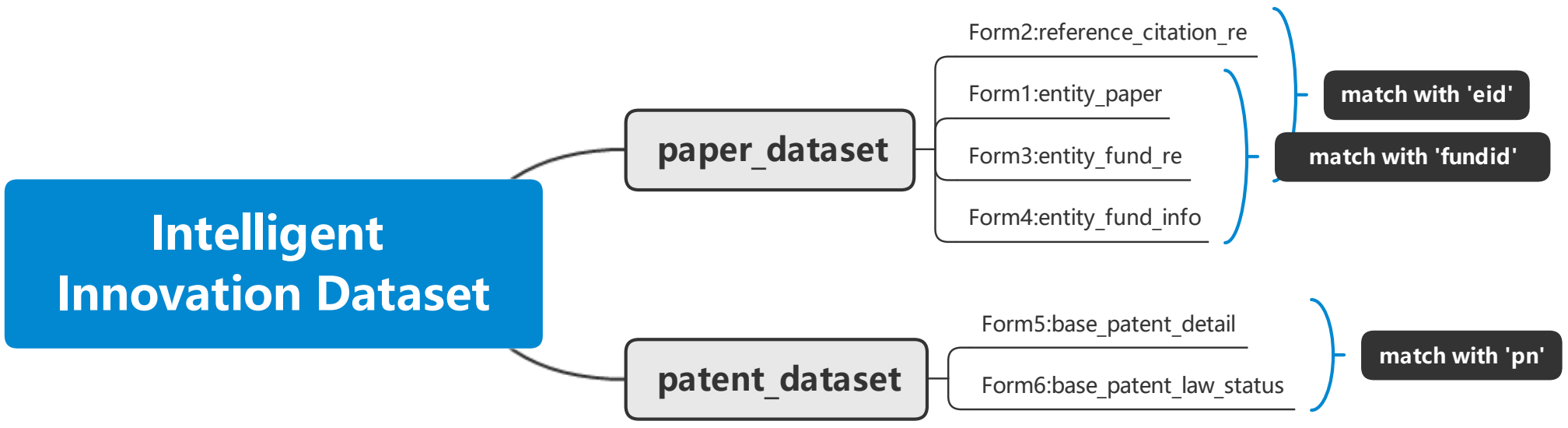

**Fig.1.** Structure of the Intelligent Innovation Dataset.

## 4 Data Processing, Integration Methods and Data Samples

### *4.1* Data Processing, Integration Methods

The collected data underwent cleaning, transformation, and other preprocessing to meet quality requirements and ensure data integration. As shown in Fig.2, the main data processing procedure includes redundant handling, missing data handling, outlier handling, data integration and data subset extraction.

(1) **Redundancy Handling:** Data fields are compared and analyzed according to standards, automatically removing duplicate fields and data.

(2) **Missing Data Handling:** Flexible completion methods address missing records due to system errors or manual entry mistakes.

(3) **Outlier Handling:** Automatic detection and treatment of data based on defined rules, allowing for customization of business rules to handle outliers, such as removal or substitution with mean values.

(4) **Data Integration:** Papers, patents, and other data are integrated to resolve issues of duplication, consistency, and uniqueness.

- Journal Papers: Data from global mainstream journals is integrated, focusing on eliminating duplication from different sources and resolving errors due to string differences.
- Conference Papers: Data from major global academic conferences is integrated to address duplication of submissions to different conferences.
- Patent Data: Global patent data is integrated to deduplicate patents filed and granted in different countries with the same priority, ensuring content consistency.
- Research Funding Information: Data from various countries' research funding information is integrated to eliminate duplication and ensure uniqueness.

(5) **Data Subset Extraction:** When using or analyzing large datasets, it may be necessary to extract specific portions, achieved through subset extraction.

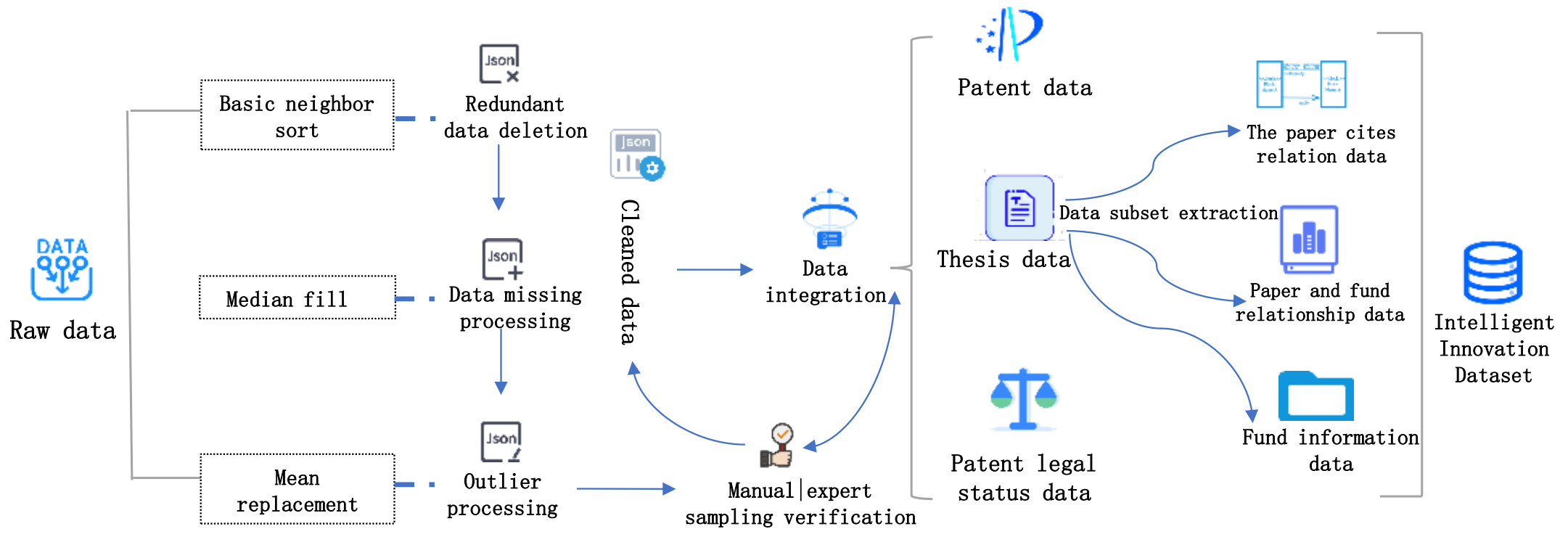

**Fig.2.** Data processing flow chart.

## *4.2* Data sample and field format

(1) Paper table and paper citation relationship table

Paper table (see Table 1) contains over 92 million paper records spanning from year 1788 to year 2023. It includes essential data such as paper titles, abstracts, journal information, author details, citation counts, and publication years.

**Table 1 entity_paper**

| Fields | Data type | Explain | Sample |
|---|---|---|---|
| eid | bigint | Get the original article id | 18679496 |
| paper_id | bigint | Unique article id in the database | 800000001 |
| source_id | bigint | Collect and obtain the original journal id | 34994 |
| source_title | text | Journal title | Journal of Sedimentary Petrology |
| paper_title | text | Article title | Sediment transport during the winter on the Yukon prodelta, Norton Sound, Alaska. |
| paper_summary | text | Abstract of article | Winter in the N.Bering Sea brings a drastic reduction in terrestrial runoff and a substan-tial decrease in wind and waves owing to the formation of shorefast and pack ice. Despite these changes, quantities of suspended silt and clay over the Yukon prodelta in the win-ter of 1978 were essentially the same as those observed during fair weather summer periods. We conclude that during the winter the suspended matter transport system is driven by tidal current reworking of sedi-ments which were introduced by the Yukon |
| | | | River during the previous summer. -from Authors. |
| paper_authors_name | json | Author name information | ["Drake D.E.", "Totman C.E.", "Wiberg P.L."] |
| paper_authors_id | json | Author id information | ["7102454169", "6508035846", "7004126924"] |
| pub_year | int | Publication year | 1979 |
| cited_num | int | Cited number | 2 |
| paper_type | varchar | Publication types | Article |
| first_author_id | text | First author | 7102454169 |
| author_keyword_json | text | Author keyword type:0 | ["Eu<sup>3+</sup>", "fluorides", "Luminescence", "quantum efficiency"] |
| index_keyword_json | text | Index keyword type:1 | ["DEFINITION OF SECURE DRIVING", "DEVELOPMENT OF VEHICLE ERGONOMICS", "MAIN CAUSES OF INJURIES AND PREMATURE DEATH", "PEDESTRIAN ENDANGERING", "REDUCTION OF TRAFFIC ACCIDENTS", "TRAFFIC DENSITY", "AUTOMOBILE DRIVERS"] |
| group_subject_id | json | Literature first-level discipline belonging | [11] |
| group_asjc | json | Literature secondary subject belonging | [2200] |
| doi | varchar | A unique document identifier | 10.1306/212F78DD-2B24-11D7-8648000102C1865D |
| language | varchar | Publication language | English |
| authors_info | json | Author list | [{"id": "7102454169", "name": "Drake D.E.", "email": null, "affiliationReferences": []}, {"id": "6508035846", "name": "Totman C.E.", "email": null, "affiliationReferences": []}, {"id": "7004126924", "name": "Wiberg P.L.", "email": null, "affiliationReferences": []}] |
| affiliations_info | json | List of affiliated institutions | [{"id": "60021199", "name": "Center of Molecular Biology CNR, Institutes of Chemistry and Biochemistry, Faculty of Medicine, Rome, Italy", "fullName": "Center of Molecular Biology CNR, Institutes of Chemistry and Biochemistry, Faculty of Medicine", "reference": "a", "fullAddress": "Rome, Italy", "departmentId": "105445006"}, {"id": "60011576", "name": "Institute of Organic Chemistry, University of Messina, Messina, Italy", "fullName": "Institute of Organic |
| | | | Chemistry, University of Messina", "reference": "b", "fullAddress": "Messina, Italy", "departmentId": "106030295"}, {"id": "60010062", "name": "Istituto Superiore di Sanità, Rome, Italy", "fullName": "Istituto Superiore di Sanità", "reference": "c", "fullAddress": "Rome, Italy", "departmentId": null}] |
| source_info | json | Journal source information | {"id": "34994", "isbn": null, "issn": "00224472", "type": "j", "coden": null, "isbns": null, "issue": "4", "pages": "1171 - 1180", "title": "Journal of Sedimentary Petrology", "volume": "49", "publisher": null, "articleNumber": null, "volumeEditors": null, "additionalInfo": null, "firstPublication": {"date": "1979", "year": "1979"}, "subjectAreaCodes": ["2300", "1900"], "conferenceSponsors": null, "additionalInfoDetails": null, "abbreviatedSourceTitle": null} |
| open_access | tinyint | Open Access: 0False, 1True | 0 |
| pubmed_info | json | PubMed information | {"id": "8619631"} |
| update_notice | json | Document update information | |
| chemicals_info | json | Chemical element information | [{"source": "esbd", "chemicals": [{"name": "hemocyanin", "casRegistryNumber": ["9013-72-3"]}, {"name": "mercuric chloride", "casRegistryNumber": ["7487-94-7"]}]}] |
| sequence_banks | json | Sequence database | [{"name": "GENBANK", "sequenceNumbers": [{"name": "AY159319", "type": "referenced"}]}] |
| author_keywords | json | Author-provided keywords | ["Antipsychotics", "Creativity", "Opiates", "Psychosis", "Ritual", "Shelley", "Spirituality"] |
| collaborations | json | Publication collaboration information | |
| funding_list | json | Funding source list | [{"acronym": "NASA", "numbers": "2206730,NNX15CC34C", "sponsor": "National Aeronautics and Space Administration", "sponsorLink": "https://www.fundinginstitutional.com/funders/100000104"}] |
| funding_detial | text | Funding summary information | This research was supported by NASA STTR Phase II under Contract no. |
| | | | NNX15CC34C and Northrop Grumman Grant no. 2206730. The authors would like to thank Jack Maceachern for his careful proofreading of this paper. |
| tradenames_info | json | Document information update | {"drugTradenames": null, "deviceTradenames": null, "textileTradenames": null} |
| correspondences | json | Corresponding author information | [{"person": "W.E. Ladd", "eaddress": null, "affiliation": "Boston, MA, United States"}] |
| indexed_keywords | json | Indexed keywords | {"CCV": ["CATALYSTS"], "CLU": ["VACUUM DISTILLATES"], "CMH": ["HYDROCARBONS"]} |
| open_access_types | json | Open access type in-formation | ["GREEN_OPEN_ACCESS", "ALL_OPEN_ACCESS", "GOLD_OPEN_ACCESS"] |
| publication_stage | varchar | Publication status | FINAL |
| order_document_link | varchar | Subscription flag | |
| view_at_repository_link | text | Open access full-text URL | http://downloads.hindawi.com/jour-nals/ijap/2018/4517848.pdf |
| tradenames_manufac-turers_info | json | Trademark infor-mation | {"drugTradenames": null, "deviceTradenames": null, "textileTradenames": null} |
| create_time | timestamp | Creation time | 2022-03-24 19:06:38 |
| update_time | timestamp | Update time | 2023-07-14 15:16:44 |

Paper citation relationship table (see Table 2) contains the citation relationships. Here, ref_id represents the reference paper, while cited_id represents the cited paper (essentially both are eid).

**Table 2 reference_citation_re**

| **Fields** | **Data type** | **Explain** | **Sample** |
|---|---|---|---|
| ref_if | bigint | References - upstream | 3552601 |
| cited_id | bigint | Citations - downstream | 3595 |

(2) Funding relationship table and funding information table

Funding relationship table (see Table 3) contains the relationships between papers and funded research projects.

**Table 3 entity_fund_re**

| **Fields** | **Data type** | **Explain** | **Sample** |
|---|---|---|---|
| eid | bigint | Get the original article id | 10711 |
| fid | bigint | Fund ID | 100007197 |
| fname | varchar | Fund Name | U.S. Pubinglic Health Service |
| fnumber | varchar | Grant Project Number. | 5T5-GM-22-08 |

Funding information table (see Table 4) contains 25,840 fund-related information entries, including funding type, creation time, funding official website, funding description, and more. It has covered funding information from 157 countries up to 2021.

**Table 4 entity_fund_info**

| Fields | Data type | Explain | Sample |
|---|---|---|---|
| fundid | bigint | Fund ID | 501100013385 |
| fundtype | varchar | Fund Type | “National-level ““Personal fund”” Corporate fund”“Provincial and ministerial-level fund”“Government fund”[1] |
| fundname | varchar | Fund Name | National Science and Technology Program during the Twelfth Five-year Plan Period |
| country | varchar | Fund Country | China |
| year_founded | int | Creation Date | 1902 |
| website | text | Fund Official Website | https://www.3m.com/ |
| description_detial | varchar | Fund Description | Business and Industry |
| info | json | Detailed Information | {"hidden": false, "country": "chn", "orgDbId": 501100013385, "contacts": [{"url": "https://iopscience.iop.org/article/10.1088/1757-899X/392/6/062128/pdf", "address": {"city": "\n ", "state": "\n ", "street": "\n ", "country": "China", "postalCode": "\n ", "attr_country": "chn"}}], "registry": {"awardDataset": {"captured": false}, "opportunityDataset": {"captured": false}}, "asjcCodes": [], "subTypeId": "govnon", "abbrevName": "NSaTP", "showFunder": true, "contextName": "National Science and Technology \nProgram of China", "countryName": "China", "relatedOrgs": [{"orgDbId": 501100002855, "relType": "partOf", "abbrevName": "MOST"}], "totalAwards": 0, "recordSource": "https://iopscience.iop.org/article/10.1088/1757-899X/392/6/062128/pdf", "relatedItems": [], "lastUpdateDate": "2023-02-17T05:30:38Z", "fundingBodyType": "gov", "revisionHistory": {"status": "update", "createdDate": {"date": "2019-03-20T16:13:39Z", "version": 0}, "revisedDate": {"date": "2023-02-17T05:30:38Z", "version": 2}}, "preferredOrgName": "National Science and Technology Program during the Twelfth Five-year Plan Period", "subTypeDescription": "Governmental |

[1] These contents are in Chinese in table.

| | | | |
|---|---|---|---|
| | | | Organizations", "totalOpportunities": 0, "funderAddressStateName": "", "funderTypeGroupDescription": "Governmental Organizations"} |
| createtime | timestamp | Creation Time | 2023-03-28 16:51:43 |
| updatetime | timestamp | Update Time | 2023-03-28 17:24:20 |

(3) Patent table and patent legal status table

Patent table (see Table 5) contains nearly 100 million patent-related data entries, including important information such as patent numbers, patent titles, applicants, application years, and more.

**Table 5 base_patent_detail**

| **Fields** | **Data type** | **Explain** | **Sample** |
|---|---|---|---|
| pn | varchar | Publication (Announcement) Number | JP2009003245A |
| title | text | Patent Title | LIQUID CRYSTAL DISPLAY AND METHOD FOR MANUFACTURING SAME |
| abs | text | Abstract | PROBLEM TO BE SOLVED: To provide a liquid crystal display in which foreign matter or the like is prevented from depositing on a backlight unit or a liquid crystal display panel in a step of attaching the liquid crystal display panel to the backlight unit, and to provide a method for manufacturing the display. SOLUTION: When a liquid crystal display panel 61 is attached to a backlight unit 62 by causing a light receiving side of the liquid crystal display panel 61 to be closer to the light emitting side of the backlight unit 62, a gap is produced since the light emitting side of the backlight unit 62 faces the light receiving side of the liquid crystal display panel 61. The light emitting side of the backlight unit 62 in contact with the gap is protected by a protection sheet 41 for a light emitting unit. Whereas, the light receiving side of the liquid crystal display panel 61 in contact with the gap is protected by a protection sheet 14a for a backside polarizing plate. COPYRIGHT: (C)2009,JPO&INPIT |
| pr | json | Priority Number | ["JP2007164960A 2007-06-22"] |
| ap_or | json | Applicant | ["SHARP KK"] |
| in_or | json | Inventor (Designer) | ["MATSUDA TAKEFUMI"] |
| ipc | json | | ["G02F1/1333", "G02F1/13357"] |
| cpc | json | | ["H01M2004/029 (EP,US)", "H01M6/48 (EP,US)", "Y02E60/10 (EP)", "H01M10/18 (EP,US)", "H01M4/14 (EP,US)"] |
| pn_date | json | Publication (Announcement) Date | ["2013-07-31"] |
| ad | json | Filing Date | ["2007-06-22"] |
| family_number | varchar | Family Number | 048852338 |
| year | int | Application Year | 2007 |

Patent legal status table (see Table 6) contains legal information for each stage of the patents.

**Table 6 base_patent_law_status**

| **Fields** | **Data type** | **Explain** | **Sample** |
|---|---|---|---|
| pn | varchar | Patent Number | WO2017052849A1 |
| event_date | varchar | Legal Status Date | 20170620 |
| authorize | Int | Grant Effective | 0 |
| reject | Int | Abandoned, Rejected, Refused, Revoked | 0 |
| event_code | varchar | Legal Status Code | PB01 |
| code_expl | varchar | Legal Status Information | PUBLICATION |
| transfer | Int | Assignment, Transfer | 0 |
| invalid | Int | Terminated, Expired | 0 |

*4.3* Data statistics distribution

We conducted some statistical analysis on the basic information in the dataset and created figures as shown below.

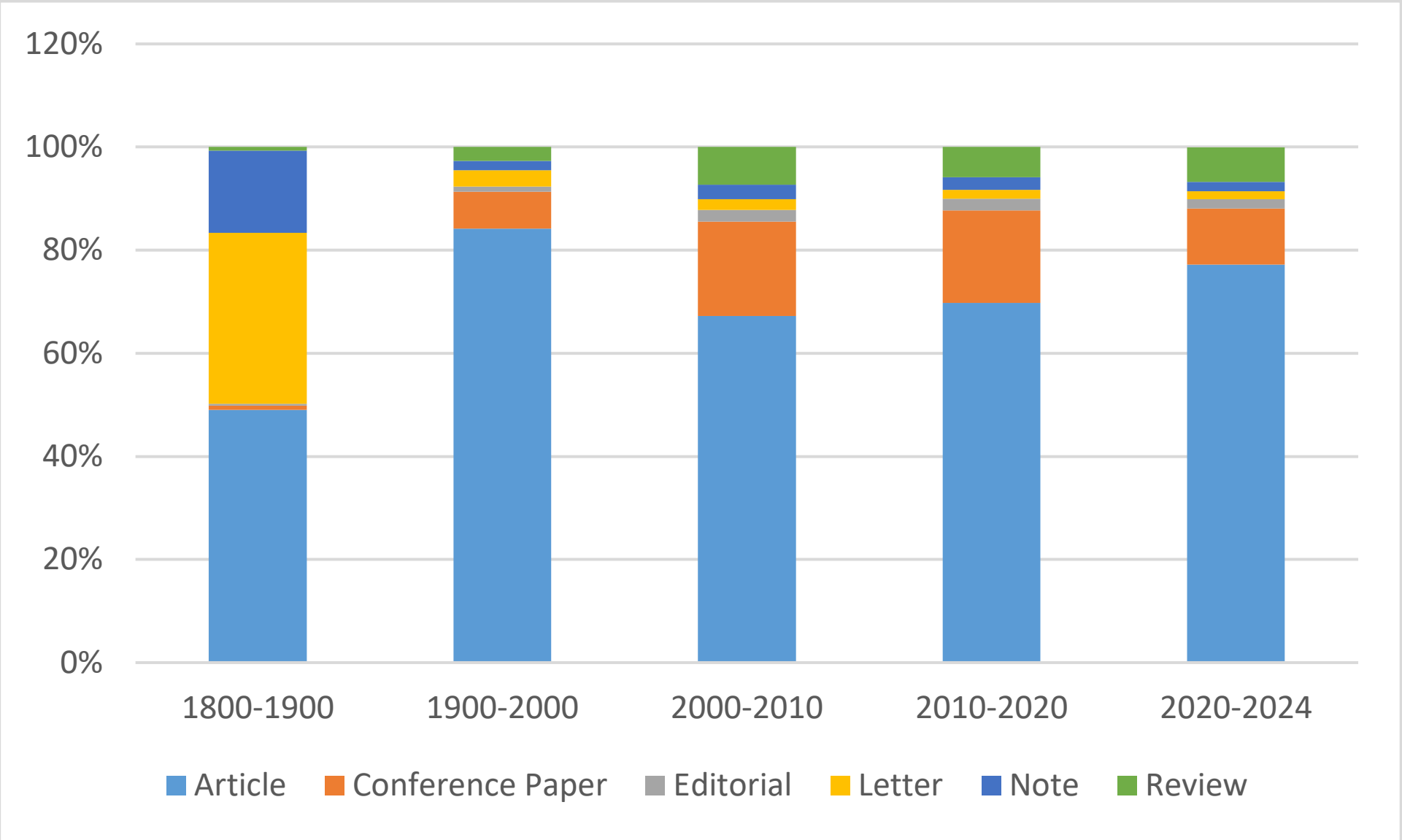

**Fig.3.** Publication types statistics—on the proportion of various types of papers in different years, including six main types: Article, Conference Paper, Editorial, Letter, Note, and Review.

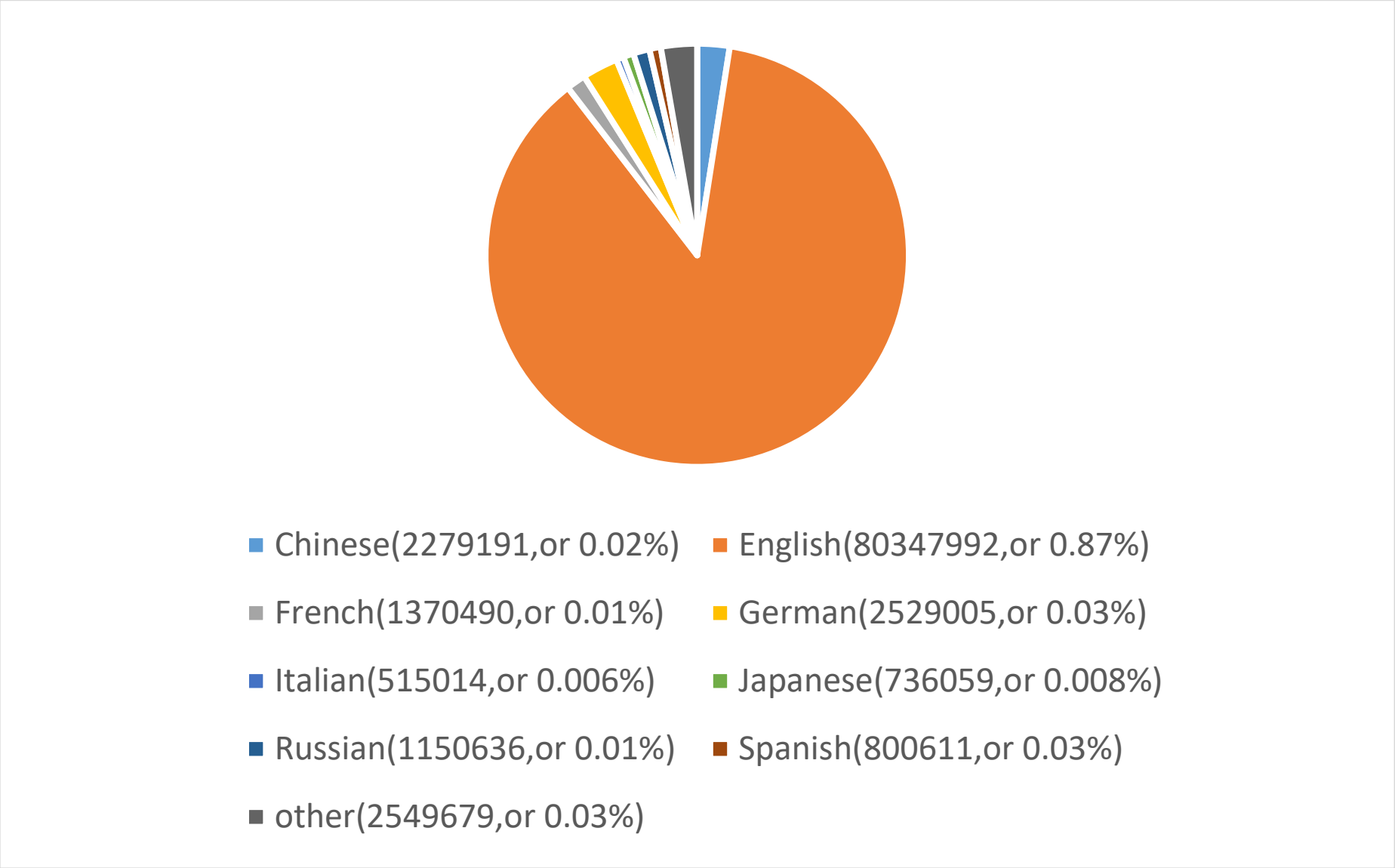

**Fig.4.** Publication language distribution. The papers in the Paper table cover 63 different languages. Papers in English, Chinese, Russian, Spanish, German, French, Japanese, and Italian account for 99% of the total volume.

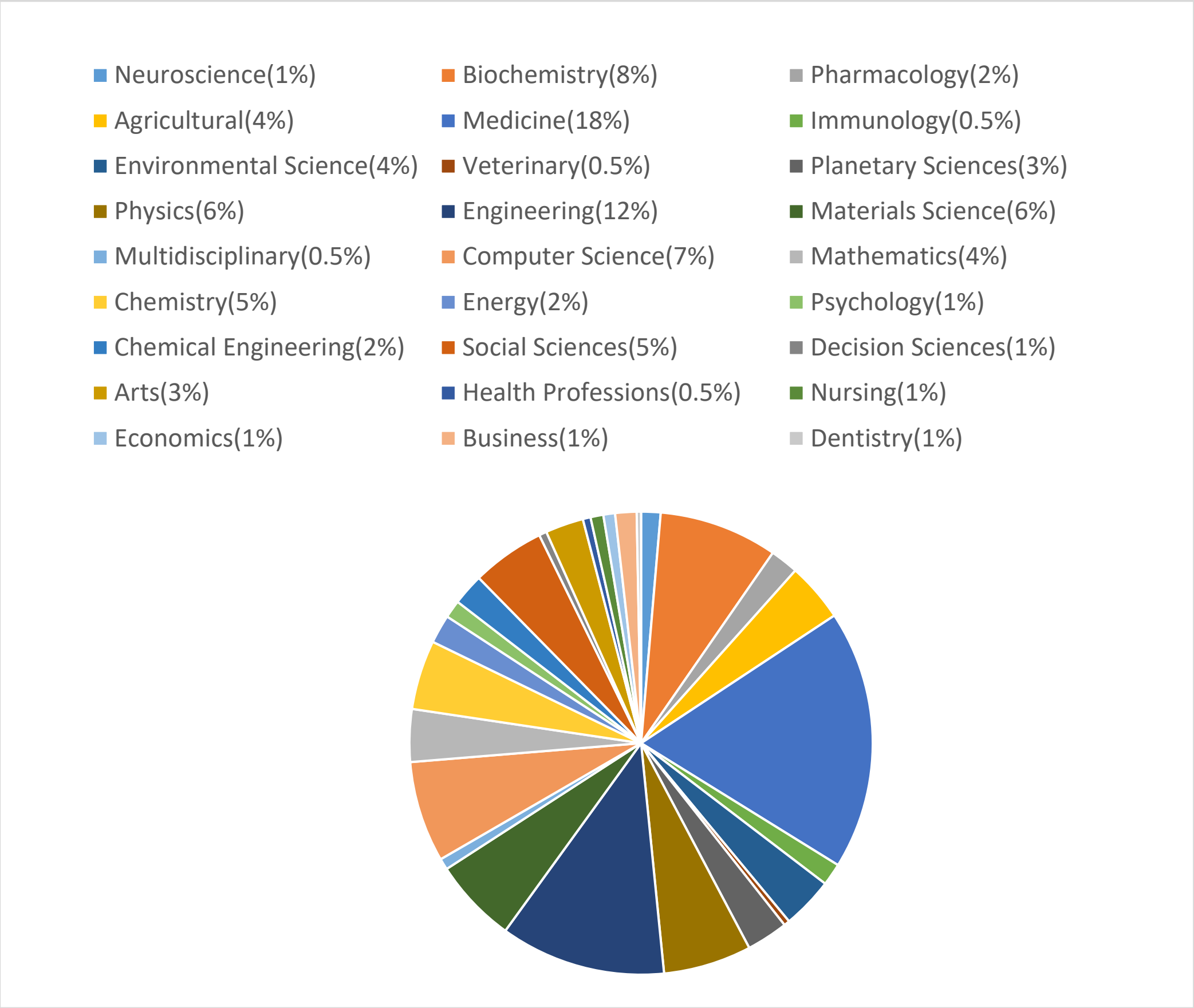

**Fig.5.** Subject distribution. The Paper table involves 27 first-level disciplines. Each paper is categorized into one or more related disciplinary fields.

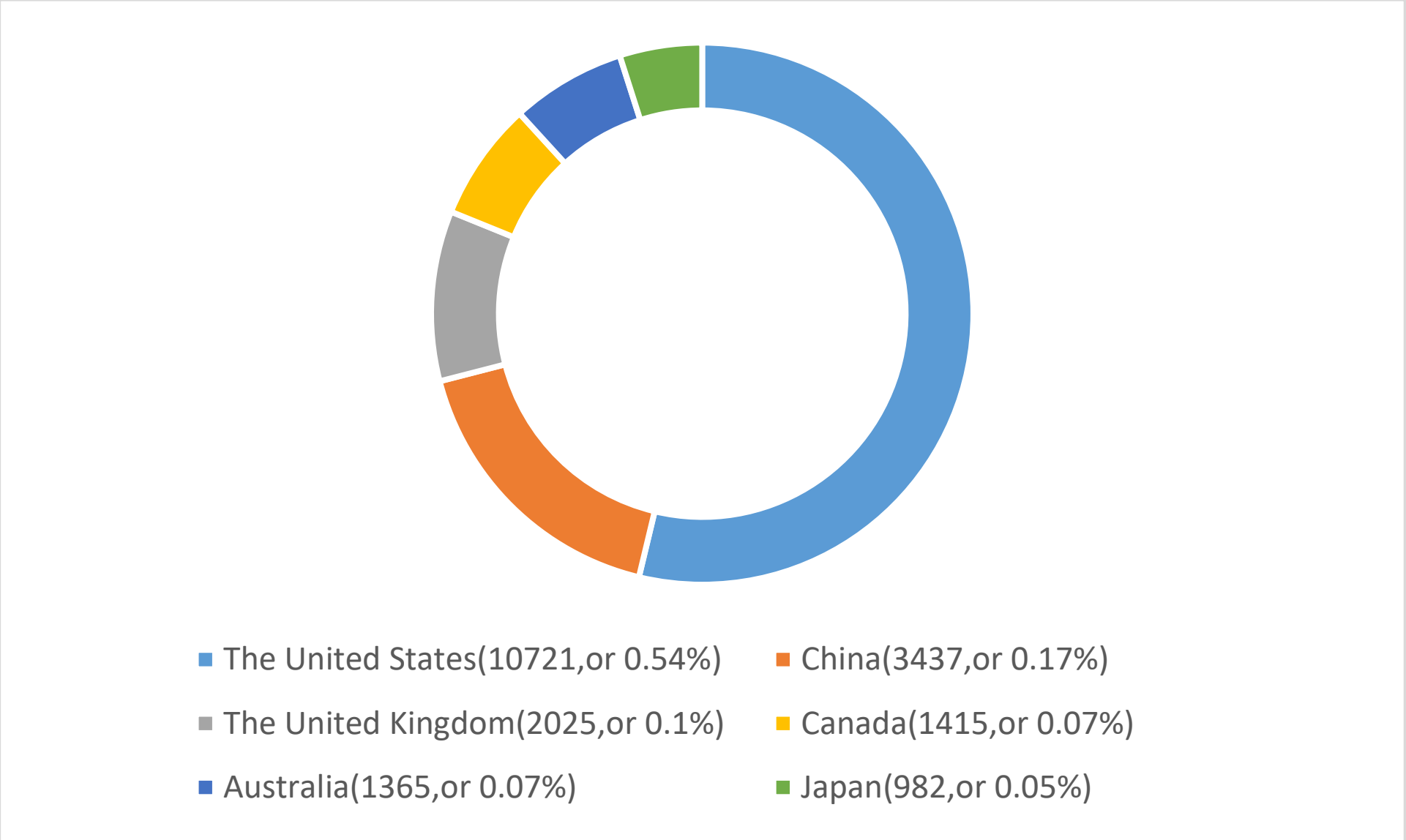

**Fig.6.** Funding quantity distribution. Distribution of funding quantities across different countries (only listing the top five): It can be seen that funds belonging to the United States account for the largest proportion, followed by China and the United Kingdom.

## 5 Comparison between IIDS and existing datasets

A comparison of main features between IIDS and two existing representative datasets, i.e., OpenAlex and Patents View datasets is given in Table 7. Below is a brief introductions to the OpenAlex and Patent View datasets and how IIDS differ from them.

OpenAlex, created by the non-profit organization OurResearch, is a fully open directory of the global research system, dedicated to building a comprehensive index of academic works, including papers, datasets, books, and more. The dataset meticulously documents academic entities and their interconnections, such as works, authors, publication sources, research institutions, research topics, publishers, and funders.

Patents View dataset encompasses a variety of records related to locations, inventors, and International Patent Classifications (IPC), totaling 2,739,749 patents originating from either the United States (US) or China (CN). Among these patents, 1,657,501 are associated with multiple inventors and multiple IPC classifications, indicating a significant level of collaborative effort. It is noteworthy that this dataset exclusively includes collaborations between Chinese and American inventors within their respective national groups.

In contrast, IIDS not only encompasses a vast array of academic work data but also includes a substantial collection of patent information, spanning a broader range of years compared to OpenAlex and Patent View Dataset.

**Table 7 Comparison among IIDS, OpenAlex[8] and Patents View datasets[9]**

| | Intelligent Innovation Dataset | OpenAlex Dataset | Patents View Dataset |
|---|---|---|---|
| Whether it includes paper data | Yes | Yes | No |
| Number of paper records | 92.31 million | 250 million | N/A |
| Year range of paper records | From 1788 to 2024 | From 1825 to 2024 | N/A |
| Number of paper types | 16 | 20 | N/A |
| Whether it includes paper language information | Yes | No | N/A |
| Whether it includes funding data | Yes | Yes | No |
| Amount of funding records | 30,000 | 32,000 | N/A |
| Whether it includes patent data | Yes | No | Yes |
| Number of patent records | 100 million | N/A | About 2.7 million |
| Source of patent data | European patent office | N/A | China and USA |
| Year range of patent records | From 1950 to 2024 | N/A | From 1980 to 2024 |

## Limitations

This study is limited in that it has not fully explored the interconnectedness between the author fields in the paper database and the patent database, preventing the full integration of patent and paper data. Addressing this issue is considered for future work.

## References

[1] Hassan, M., A. Zaveri, and J. Lehmann. “A linked open data representation of patents registered in the US from 2005–2017.” *Sci Data* 5(2018):180279. https://doi.org/10.1038/sdata.2018.279

[2] de Rassenfosse, G. and F. Seliger. “Imputation of missing information in worldwide patent data.” Data in Brief 34(2021):106615. https://doi.org/10.1016/j.dib.2020.106615.

[3] Mogee, M. E. “Using Patent Data for Technology Analysis and Planning.” *Research-Technology Management34.4*(1991): 43–49. https://doi.org/10.1080/08956308.1991.11670755

[4] A.M. Phirouzabadi, D. Savage, J. Juniper, K. Blackmore. “Dataset on the global patent networks within and between vehicle powertrain technologies—cases of ICEV, HEV, and BEV.” Data in Brief 28(2020):105017. https://doi.org/10.1016/j.dib.2019.105017.

[5] Carpenter, M. P., F. Narin, P. Woolf. “Citation rates to technologically important patents.” World Patent Information 3.4 (1981): 160-163. https://doi.org/10.1016/0172-2190(81)90098-3.

[6] Niebel, T., F. Rasel, S. Viete. “BIG data – BIG gains? Understanding the link between big data analytics and innovation.” *Economics of Innovation and New Technology*, *28*.3(2018): 296–316. https://doi.org/10.1080/10438599.2018.1493075

[7] Pine, K. H., M. Hinrichs, K. Love, M. Shafer, G. Runger, and W. Riley. "Addressing Fragmentation of Health Services through Data-Driven Knowledge Co-Production within a Boundary Organization." The Journal of Community Informatics 18.2 (2022): 3-26. https://doi.org/ 10.15353/joci.v18i2.3595.

[8] Priem, J. ,H. Piwowar, and R. Orr. "OpenAlex: A fully-open index of scholarly works, authors, venues, institutions, and concepts." arXiv e-prints arXiv:2205.01833 (2022). https://arxiv.org/abs/2205.01833 (accessed on September 20, 2024).

[9] Patents View, https://github.com/PatentsView/ (accessed on September 20, 2024).